%% file: main.tex
\documentclass[sigconf]{acmart}

\renewcommand\footnotetextcopyrightpermission[1]{}

\usepackage{booktabs}
\usepackage{multirow}
\usepackage{enumitem}
\usepackage{xspace}
\usepackage{dblfloatfix}
\usepackage{xcolor,colortbl}
\usepackage{hhline}
\usepackage{makecell}
\usepackage{placeins}

\newcommand{\MyBGColorA}{{\cellcolor{gray!15}}}
\newcommand{\MyBGColorB}{{\cellcolor{gray!30}}}
\newcommand{\MyBGColorC}{{\cellcolor{gray!45}}}

\usepackage{caption}
\setcopyright{rightsretained}

\acmConference[]{}{}{} 
\acmYear{2018}
\copyrightyear{2018}

\acmPrice{15.00}

\acmSubmissionID{123-A12-B3}

\begin{document}

\title{Forecasting Suspicious Account Activity at\\Large-Scale Online Service Providers}

\author{Hassan Halawa}
\affiliation{%
  \institution{University of British Columbia}
  \city{Vancouver} 
  \state{British Columbia} 
  \country{Canada}
}
\email{hhalawa@ece.ubc.ca}

\author{Matei Ripeanu}
\affiliation{%
  \institution{University of British Columbia}
  \city{Vancouver} 
  \state{British Columbia} 
  \country{Canada}
}
\email{matei@ece.ubc.ca}

\author{Konstantin Beznosov}
\affiliation{%
  \institution{University of British Columbia}
  \city{Vancouver} 
  \state{British Columbia} 
  \country{Canada}
}
\email{beznosov@ece.ubc.ca}

\author{Baris Coskun}
\affiliation{
  \institution{Amazon AI}
  \city{New York} 
  \country{USA}
}
\email{barisco@amazon.com}

\author{Meizhu Liu}
\affiliation{%
  \institution{Yahoo Research}
  \city{New York}
  \country{USA} 
}
\email{meizhu@oath.com}

\renewcommand{\shorttitle}{}

\newcommand{\companylongname}{\textit{Large-Scale Online Service Provider}\xspace}
\newcommand{\companyshortname}{\textit{LSOSP}\xspace}

\begin{abstract}
In the face of large-scale automated social engineering attacks to large online services, fast detection and remediation of compromised accounts are crucial to limit the spread of new attacks and to mitigate the overall damage to users, companies, and the public at large. We advocate a fully automated approach based on machine learning: we develop an early warning system that harnesses account activity traces to predict which accounts are likely to be compromised in the future and generate suspicious activity. We hypothesize that this early warning is key for a more timely detection of compromised accounts and consequently faster remediation. We demonstrate the feasibility and applicability of the system through an experiment at a large-scale online service provider using four months of real-world production data encompassing hundreds of millions of users. We show that---even using only login data to derive features with low computational cost, and a basic model selection approach---our classifier can be tuned to achieve good classification precision when used for forecasting. Our system correctly identifies \textit{up to one month in advance} the accounts later flagged as suspicious with precision, recall, and false positive rates that indicate the mechanism is likely to prove valuable in operational settings to support additional layers of defense. 
\end{abstract}

\maketitle

\input{body}

\input{main.bbl}

\end{document}

%% file: body.tex
\section{Introduction}
\label{section:introduction}

Online services are an integral part of our personal and professional lives. To support widespread adoption and improve usability, large-scale online service providers (LSOSPs) have made it simple for users to access any of the provided services using a single credential. Such ``single sign-on'' systems make it much easier for users to manage their interactions through a single account and sign-in interface. As users become more invested in the platform, the single login credential becomes a valuable key to a whole set of services, as well as the `key' to their digital identity and the personal information stored on the platform. As a consequence, these credentials are highly attractive targets to attackers as well.

As LSOSPs improve their defense systems to protect their user base, attackers have shifted their efforts to social engineering attacks, e.g, attacks that exploit incorrect decisions made by individual users to trick them into disclosing their login credentials~\cite{jagatic:2007}. Once an account is compromised, the attackers hijack the account from its legitimate owner and, typically, use it for their own purposes~\cite{moore:2009}: for example, to evade detection while perpetuating an attack (e.g., multi-stage phishing, or malware distribution campaigns) or to carry out other fraudulent activity (e.g., sending out spam email).

Thus, detecting compromised accounts early and giving back control to their legitimate owners quickly, and even designing defense mechanisms that add additional layers of defense to protect users likely to fall prey to social engineering attacks, is crucial. Doing so can mitigate the damage an attacker can do while in control of a compromised account, protect the account owner's digital identity, and reduce the damage inflicted by an automated large-scale social-engineering attack to a LSOSP and its user community. It should be noted that, detecting compromised accounts is much more challenging than just identifying fake ones (i.e., those created by an attacker) since, in such cases, suspicious activity is typically interleaved with the account owner's legitimate activity~\cite{egele:2013}.

This paper tests the hypothesis that it is feasible to identify likely future victims of mass-scale social-engineering attacks. In a nutshell, we postulate that the behavioral patterns of the users that have little incentives or low ability to fend off social-engineering attacks can be learned. To this end we propose an early warning system based on a completely automated pipeline using machine learning (ML) to identify the accounts with similar behavioral patterns to accounts that have been flagged as suspicious in the past.  

We postulate that predicting accounts that are more likely to be compromised in the future can be used to develop new defenses, to fine-tune and better target existing defense mechanisms, as well as to better protect vulnerable users~\cite{halawa:2016}.
For example, in the context of online social networks, identifying potentially vulnerable accounts - even with low accuracy - was sufficient to develop a defense system for detecting fake accounts that significantly outperformed other state of the art approaches~\cite{boshmaf:2015}.
While we briefly discuss the intuition behind some of these defense mechanisms in the discussion section (\S\ref{section:discussion}), their design and evaluation, however, is beyond the scope of this paper and we focus here solely on evaluating our conjecture that predicting accounts which are more likely to be compromised is feasible.

\begin{figure*}
    \setlength{\belowcaptionskip}{-12pt}
    \includegraphics[width=0.75\textwidth,height=1.25in,keepaspectratio]{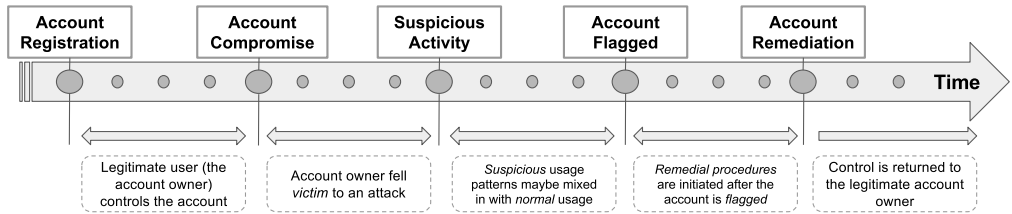}
    \caption{Overview of the Life Cycle of a Compromised Account}
    \label{figure:account-life-cycle-overview}
\end{figure*}

We have tested our hypothesis using real-world data from a large LSOSP (i.e., at the scale of Amazon, Facebook, Google, Yahoo, etc.). Throughout this paper we will refer to it as a 
\textit{\companyshortname} (in italics, the non-italicized LSOSP refers to a generic Large-Scale Online Service Provider).
Our experiments were carried out over four months worth of production data covering hundreds of millions of users generating hundreds of billions of login events to \textit{\companyshortname}'s platform.

Overall, our results indicate that it is feasible to achieve good classification accuracy as well as a low false positive rate. Our evaluation demonstrates that the proposed approach is not only feasible but also offers promising performance, based on which new or improved defense mechanisms can be developed. It is important to note that our results should be seen as a lower bound of achievable classification performance, which can likely be further improved by using richer data or additional computational resources (e.g., to support more sophisticated learning methods) as we discuss in \S\ref{section:discussion}.

This paper makes the following contributions:
\begin{itemize}[noitemsep,topsep=0pt,leftmargin=*]
    \renewcommand{\labelitemi}{\scriptsize$\blacksquare$}

    \item We formulate the hypothesis that it is feasible to identify the users more likely to fall prey to mass-scale social-engineering attacks (\S\ref{section:problem-formulation}), propose an approach to identify these accounts, and outline the design of such an early warning system (\S\ref{section:proposed-methodology}).
    
    \item We demonstrate the feasibility and applicability of the proposed approach on real-world production data (\S\ref{section:results}). We show that, even using low-cost features extracted from two basic datasets (\S\ref{section:datasets}) and a simple model selection approach (\S\ref{section:proposed-methodology}) for acceptable training runtime, the proposed classifier can be tuned to achieve good classification quality based on recall, precision, and false positive rate metrics (\S\ref{section:results}). For example (\S\ref{section:results-CE-C}), using only \textit{one week of login event history and predicting one month in advance}, our classifier predicts more than half of the accounts later flagged as having suspicious behaviour (i.e., achieves a recall of 50.62\%) and, at the same time, around one in five of the accounts predicted to generate suspicious behaviour is actually labeled as suspicious at \textit{\companyshortname} within a 30-day prediction horizon (i.e., precision of 18.33\%, with a corresponding false positive rate of 0.49\%).

\end{itemize}

\section{Background: Account Life Cycle}
\label{section:background}

Figure~\ref{figure:account-life-cycle-overview} provides an overview of the typical life cycle of an account that at some point is \textit{compromised} (e.g., through a social engineering attack). After registering an account, a legitimate user starts using the provided services. We refer to the behavioral patterns over this time as the user's \textit{normal} behavior. Attackers may carry out automated social engineering attacks that attempt to compromise user accounts (e.g., falling for a phishing attack)~\cite{jagatic:2007, boshmaf:2011, yan:2011}.

If the attack is successful, the user becomes a \textit{victim} of the attack and her account is considered compromised, and eventually the attacker may start to (mis)use the account. Typically, some time is required before the defense systems employed by LSOSPs can identify such activity and, in turn, \textit{flag} the account as \textit{suspicious}.

At that stage, \textit{remedial procedures} at the LSOSP start to take place~\cite{freeman:2016}. For example, a popular low-cost measure is to use CAPTCHAs to make it more difficult for an attacker to automate the use of compromised accounts.  The protective measures can be escalated and extra challenges can be then thrown at identified suspicious accounts such as prompting for answers to secret questions, or second factor authentication (2FA) through another service or through a mobile device. Their goal is to prevent the attacker from continuing to use the account while providing the legitimate user with some means of regaining control. After remediation, the user takes back control of her account. The account is no longer compromised and the observed behavioral patterns return back to normal over time.

If the suspicious behavior persists, the employed protective measures can be escalated again to limit potential damage to other users on the platform. For example, the LSOSP can quarantine the suspicious account (e.g., impose rate limits on the outbound messages or deny certain types of service) or even suspend it.

We note that \textit{fake} accounts, i.e., accounts created by the attacker from the beginning, share some of the same phases during their lifetime as those discussed above for legitimate accounts. Fake accounts are typically disguised to appear legitimate in order to evade detection for as long as possible. For example, realistic looking profiles can be faked, normal behavioral patterns can be emulated, and fake accounts can be registered through unique IP addresses to avoid clustering.

\section{Problem Formulation}
\label{section:problem-formulation}

We present a formal description of our problem by abstracting away from all company- and experiment-specific details. We describe those details in \S\ref{section:proposed-methodology},  \S\ref{section:datasets} and \S\ref{section:results} respectively. Here, we go over the assumptions and objectives that influenced our approach, we elaborate on the datasets required to carry out the classification task (\S\ref{section:assumptions}), and we introduce our \textit{classification exercises  (CEs)}, which are the means by which we organize our \textit{experiments} (\S\ref{section:experiment-organization}).

\subsection{Overview}

\textit{Our goal is to develop an early warning system that can be used by LSOSPs to harness observable legitimate user behavior to identify accounts likely to be labeled as suspicious in the future}. Our intuition is the following: over the course of everyday use, the history of user interactions encapsulates information from which one can infer whether an account is more likely to be compromised (e.g., because the user does not have the interest or the ability to fend off social-engineering attacks); eventually (some of) these accounts are compromised, generate suspicious activity, and may later be flagged. 
In other words, to forecast future suspicious activity, we aim for features that approximate user behavioral patterns to infer similarity to legitimate accounts that are later flagged as suspicious.

We are developing a binary classifier to act as an \textit{early warning system}.  We chose a supervised machine learning classifier as, over the past few years, such approaches have been shown to achieve good performance for a variety of classification tasks~\cite{liu:2015, soska:2014, shon:2007}.

\subsection{Assumptions, Objectives, and Datasets}
\label{section:assumptions}

\textit{\textbf{Assumptions.}} We treat the prediction of suspicious accounts as a binary classification problem (suspicious vs. non-suspicious). We assume that only a small subset of the overall population is likely to exhibit suspicious activity. We believe that this is true for large providers that offer services to a large number of users around the world (up to billions of users) and dedicate resources to maintain a `healthy' user population. The direct implication is that the ML techniques used, the data selection for the training of the classifiers, and the success metrics used are all tuned for imbalanced data.

\vspace{+3pt}
\noindent
\textit{\textbf{Objectives}}. 
We aim to meet the following objectives when designing and tuning the binary classifier. First, a low rate of false positives: accounts incorrectly predicted as suspicious (i.e., false positives) should be minimized even at the cost of decreasing the number of correctly predicted suspicious accounts (i.e., true positives). This trade-off can be controlled by tuning the classifier's prediction threshold when generating the final binary classification. We also discuss tuning for a low rate of false negatives in (\S\ref{section:discussion}).

Second, and crucially for deployment at a LSOSP, the classifier should be optimized for runtime efficiency during both training (feature extraction and model building) and testing/use (prediction and classification). This can be accomplished by employing features that can be easily extracted/computed from the raw data, and by choosing ML models that offer a good trade-off between the quality of prediction and performance. Balancing this trade-off is crucial for timely forecasting of suspicious activity and thus faster remediation (as well as adoption in realistic settings).

\vspace{+3pt}
\noindent
\textit{\textbf{Required Datasets.}}
We assume that the LSOSP has access to at least two types of data. First, data that can be mined to extract behavioral patterns.
Second, a sample of accounts previously flagged as suspicious is required as ground truth. We detail the data we use from \textit{\companyshortname} in \S\ref{section:datasets}.

\subsection{Experiment Organization}
\label{section:experiment-organization}

Here we establish the terminology we use for the rest of this paper. We define the means by which we organize our experiments (\textit{Classification Exercises}) and we detail the categories of accounts that can be observed in the datasets and how we use them. 

\begin{figure*}
    \setlength{\belowcaptionskip}{-6pt}
    \includegraphics[width=0.7\textwidth,height=1in,keepaspectratio]{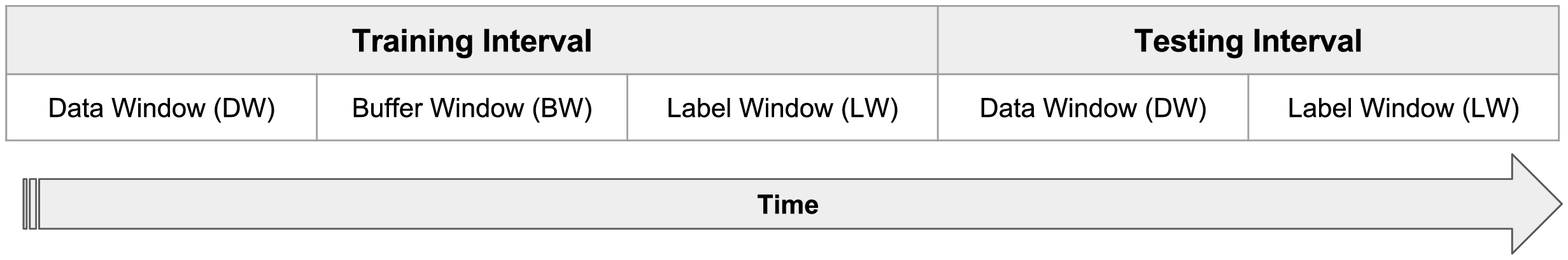}
    \caption{Overview of a Classification Exercise (CE). \textmd{Each exercise is divided into two broad phases: \textit{training}, during which the classifier is fitted, and \textit{testing}, during which the classifier predictions are evaluated. Each phase is subdivided into smaller non-overlapping time windows: \textit{Data Window} (DW), \textit{Buffer Window} (BW) and \textit{Label Window} (LW). The DW is the period of time over which behavioural features are mined. The BW is a period of time introduced to avoid learning from accounts that may already be compromised but not yet labeled as such. The LW is the period over which labels are extracted.}}
    \label{figure:classification-exercise-overview}
\end{figure*}

\vspace{+3pt}
A \textbf{Classification Exercise} (CE) is our way of grouping together all parameters of a binary classification experiment (e.g., training time interval, testing time interval, ML model hyperparameters) and the associated results. As with any typical ML approach, a CE is divided into two distinct phases: \textit{training} and \textit{testing} (Figure.~\ref{figure:classification-exercise-overview} provides an overview). During training, our goal is to fit a model that learns user behavioral patterns that can be used as early predictors of suspicious account activity. During testing, the fitted model is applied to new data \textit{not seen during training} and the classifier's performance is evaluated against a labeled ground truth.

\begin{table}
    \small
    \centering
    \caption{Set Notation Summary}
    \label{table:summary-set-notation}
    \begin{tabular}{|c|c|}

    \hline
    \textbf{Symbol} &\textbf{Description} \\
    \hline
    
    U  & Set of all registered accounts\\
    \hline
    
    $L_d$ & Set of accounts with login activity on day \textit{d} \\
    \hline
    
    $S_d$ & Set of accounts flagged as suspicious on day \textit{d} \\
    \hline

    $A_d$ & Set of accounts under attacker control on day \textit{d} \\
    \hline
    
    $F_d$ & Set of fake accounts on day \textit{d} \\
    \hline
    
    $C_d$ & Set of compromised accounts on day \textit{d} \\
    \hline
    
    $LS_{CE}$ & Set of accounts labeled as suspicious during a $CE$ \\
    \hline
    
    $P_{CE}$ & Set of accounts predicted as suspicious during a $CE$ \\
    \hline
    \end{tabular}
    \vspace{-2em}
\end{table}

\vspace{+3pt}
\noindent
\textit{\textbf{Categories of Accounts.}} Table~\ref{table:summary-set-notation} presents a summary of the set notation used to categorize user accounts. We consider $U$ as the set of all accounts registered with the LSOSP. Depending on the scale and popularity of the LSOSP, $U$ can be extremely large potentially exceeding a billion users. 

We use days as a coarse-grain measure of time. We consider $L_d$ as the set of users with login activity on day \textit{d}. For the set $L_d$, we extract easy-to-compute low-cost features representing the users' login behavior on day \textit{d}.
We aim to learn the behavioral patterns of legitimate accounts prior to them being flagged as suspicious and, as such, require some information about accounts later flagged as having suspicious behaviour. We denote with $S_d$ the set of user accounts flagged as suspicious on day \textit{d}. 
Existence of an account in set $S_d$ on day \textit{d} is a clear indication that the account exhibited some suspicious activity prior to or on day \textit{d}.

However, it is important to note that the opposite is not true: if an account is absent from the set $S_d$ on day \textit{d} that does not imply that it did not exhibit any irregular activity prior to or on day \textit{d}. The reason for this is that the pipeline used for detecting suspicious accounts at the LSOSP is expected to have some lag. In other words, it takes time for an account to be flagged as suspicious after it first starts exhibiting irregular behavior.

\vspace{+3pt}
\noindent
\textit{\textbf{Avoiding Attacker-Controlled Accounts.}} The set $L_d$ contains not only legitimate user accounts but also those that are under the control of an attacker (the set $A_d$). These include fake as well as compromised accounts (considered as sets $F_d$ and $C_d$ respectively). \textit{We implement several heuristics to prune such accounts and avoid learning user behavioral patterns from accounts that may be under attacker control}.
We discuss this preprocessing step in detail in \S\ref{section:heuristics}.

\vspace{+3pt}
\noindent
\textit{\textbf{Classification exercise data.}} This data cleaning leads to $LS_{CE}$, the set of labeled accounts during our CE (where \textit{$LS_{CE}$ $\subseteq$ $S_d$ $\{$ d $|$ d $\in$ Training Interval $\}$}).
We denote the set of accounts our pipeline forecasts as suspicious during a CE as $P_{CE}$.  To evaluate the quality of our predictions, we then compare the accounts in $P_{CE}$ to the ground truth in the testing interval (\textit{$S_d$ $\{$ d $|$ d $\in$ Testing Interval $\}$}).

\vspace{+3pt}
\noindent
\textit{\textbf{Formalization.}} During training, a CE uses as input both \textit{$L_d$} and \textit{$S_d$ $\{$ d $|$ d $\in$ Training Interval $\}$} with some pre-processing to avoid learning from attacker controlled accounts. The output after training is the fitted model $M$. The set of labels used as ground truth when fitting the model is \textit{$LS_{CE}$ $\approx$ $S_d$ $-$ $(F_d \cup C_d )$ $\{$ d $|$ d $\in$ Training Interval $\}$}. During testing, the inputs are the fitted model $M$ and \textit{$L_d$ $\{$ d $|$ d $\in$ Testing Interval $\}$}. The output of the CE is the set of predicted suspicious accounts $P_{CE}$ that are evaluated against a labeled ground truth from \textit{$S_d$ $\{$ d $|$ d $\in$ Testing Interval $\}$ }.

\vspace{+3pt}
\noindent
\textit{\textbf{Success Metrics.}} Based on our formalization above, the outcome of the binary classifier can only fall in one of four possible categories: True Positive $TP = P_{CE} \cap S_d$, False Positive $FP = P_{CE} - S_d$, True Negative $TN = (P_{CE} \cup S_d)^\complement$, and False Negative $FN = S_d - P_{CE}$, where \textit{$\{$ d $|$ d $\in$ Testing Interval $\}$}.  Based on these one can derive classification precision, recall, accuracy and false positive rate. 

\FloatBarrier

\section{Proposed Approach}
\label{section:proposed-methodology}

This section outlines our proposed approach: the details of our classification exercises (\S\ref{section:proposed-methodology-ce-composition}), the proposed ML pipeline (\S\ref{section:proposed-methodology-early-warning-pipeline}), and the heuristics we implement to avoid learning from accounts under the control of an attacker and to reduce bias when evaluating our approach (\S\ref{section:heuristics}).  The following sections describe our datasets (\S\ref{section:datasets}) and the evaluation results (\S\ref{section:results}).

\subsection{Classification Exercise Composition}
\label{section:proposed-methodology-ce-composition}

We organize our classification exercises (CEs) as outlined in Figure~\ref{figure:classification-exercise-overview}. During \textit{training}, we attempt to fit a model ($M$) that learns which behavioral patterns during the training Data Window (training-DW\footnote{Where the context makes the notation unambiguous, we skip the prefix and use \textit{DW} only for \textit{training-DW} or \textit{testing-DW}.  Similarly for \textit{LW}.}) correlated to the account being labeled as suspicious later in the Label Window (LW). We introduce a Buffer Window (BW) between the DW and LW, to account for any lag (delay) in the suspicious account flagging pipeline used to generate the ground truth of suspicious accounts. The reason is that, in the absence of the BW, a lag in the pipeline will cause the fitted model to learn user behavioral patterns from accounts that are already under the control of an attacker. In \S\ref{section:heuristics}, we present our heuristics to estimate the width of the Buffer Window (BW).

During \textit{testing}, the fitted model ($M$) obtained during training, is applied during the \textit{testing-DW} to forecast the set of accounts that are likely to have suspicious behaviour ($P_{CE}$). The quality of those predictions is then evaluated against the ground truth of labeled suspicious accounts extracted from the \textit{testing-LW}.

\subsection{The Early Warning Pipeline}
\label{section:proposed-methodology-early-warning-pipeline}

\begin{figure}
    \setlength{\belowcaptionskip}{-10pt}
    \includegraphics[width=\linewidth,keepaspectratio]{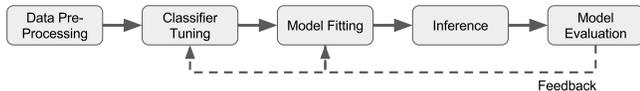}
    \caption{Overview of the Machine Learning Pipeline}
    \label{figure:pipeline-overview}
\end{figure}

Our system is composed of a pipeline that can be easily integrated into existing systems (outlined in Figure~\ref{figure:pipeline-overview}).  We note that our pipeline design stresses efficiency, scalability, and, ultimately, achieving a practical training runtime sometimes even to the detriment of the learned classifiers (e.g., using simple low-cost features as opposed to sophisticated feature extraction). With production data, similar in scale to what we have access to at \textit{\companyshortname}, our pipeline is designed to extract behavioral patterns and to train in reasonable time on log traces from hundreds of millions of accounts leading to hundreds of billions of log entries over the duration of each CE.

We developed our pipeline in Scala 2.11, employed SparkML for all our developed classifiers, and ran our CEs on Spark 2.0.2~\cite{zaharia:2010}.

\vspace{+3pt}
\noindent
\textit{\textbf{Data Pre-Processing.}}
We pre-process the datasets from which we extract the user behavioral patterns (e.g., login activity dataset) as well as the ground truth (e.g., accounts flagged as suspicious). Importantly, \textit{we also carry out a series of pruning operations in order to exclude accounts that may bias either learning or evaluation} as discussed in \S\ref{section:heuristics}.
During this stage, for each account, we extract features at the day level and aggregate them for the intervals associated with the classification exercise. There is an inherent tradeoff here: extracting and computing a large number of features over a long duration of time could potentially include more behavioral information thereby increasing the prediction accuracy. However, this comes at the cost of longer run-time and might affect prediction timeliness. At \textit{\companyshortname}, we find that extracting only a relatively small set of low-cost features that are both simple and quick enough to compute is both sufficient and also more practical from a performance perspective in a production environment (details in~\S\ref{section:datasets}). 

\vspace{+3pt}
\textit{Pre-processing Imbalanced Data.} Typically at LSOSPs, suspicious accounts (the positive class) are a minority compared to the overall population. Naively training an ML classifier on such imbalanced data will typically result in a classifier that always predicts the dominant class (the negative class in our case) to achieve the highest accuracy~\cite{provost:2001}.
Approaches to mitigate this problem include simple preprocessing techniques such as undersampling the majority class or oversampling the minority class~\cite{he:2009}, or Cost-Sensitive Learning~\cite{lomax:2013} that attempts to minimize the cost of misclassifications by assigning asymmetrical costs during the training process. At \companyshortname, given the scale of the data and our focus on building a practical pipeline with good balance between runtime and classification performance, we use undersampling during training (however, we test on the whole set of labeled data in the test set).

\vspace{+3pt}
\noindent
\textit{\textbf{Classifier Tuning.}}
Second, during the hyperparameter optimization stage, model selection is carried out in order to find the best model (or set of parameters) for the classification task. This only needs to be done once during training (or periodically, with low frequency and offline, to learn new user behavioral patterns) and is not carried out during inference using the fitted model in production. We use a Random Forest (RF) classifier considering the good trade-off it offers between runtime and classification accuracy~\cite{fernandez-delgado:2014}. We carry out the hyperparameter optimization on an independent dataset extracted from the available history and specifically reserved for this purpose (\S\ref{section:results-CE-A}). The extracted model parameters are then fixed for all the subsequent CEs.

\vspace{+3pt}
\noindent
\textit{\textbf{Model Fitting and Inference.}} 
Third, after data preprocessing and hyperparameter tuning, a ML model $M$ is fitted and later applied to make predictions on new data (i.e., inference). On the one hand, this data could be one for which there already exists labeled ground truth. In that case, the goal is to evaluate the performance of the developed classifier. On the other hand, this could be new data from production for which no ground truth exists (i.e., during the real-world deployment) and in this case, the goal is to put the classifier into practice to predict accounts likely to generate suspicious activity in the future based on their recent behavioral patterns.

\vspace{+3pt}
\noindent
\textit{\textbf{Model Evaluation.}}
Finally, we carry out model evaluation as the last stage in our pipeline. 
We obtain the confusion matrix comparing the ground truth with the resulting predictions and collect statistical measures of the classifier's performance.

\subsection{Heuristics} 
\label{section:heuristics}

Our goal is to learn behaviour from legitimate accounts (i.e., that are not attacker-controlled: fake and compromised accounts --- \textit{$A_d$ $\{$ d $|$ d $\in$ Training Interval $\}$}) and predict which legitimate accounts may later get compromised and become labeled as suspicious. To this end we use a number of heuristics. We also implement additional heuristics to increase the confidence in our evaluation results. 

\vspace{+3pt}
\noindent
\textbf{\textit{Heuristics to increase the chance that we capture only the behaviour of accounts under the control of legitimate users.}}
During training, we attempt to exclude all accounts that are potentially under the control of an attacker.  In practice, the set of accounts $A_d$ is unknown, even for historical data for which there is collected ground truth, as this set may include not-yet-detected fakes and compromised accounts. We take advantage of the fact that we have an extremely large dataset to carry out aggressive exclusions that reduce the chance that we capture behaviour from attacker-controlled accounts. We use three heuristics: First, we exclude any account flagged as suspicious during the training DW or at a later point of time within the Buffer Window (BW). By excluding these accounts, we reduce the likelihood that our classifier learns behavioral patterns stemming from detected compromised accounts. Second, to the same end, for the classification exercises where there is available data before the start of the training interval (\S\ref{section:results-CE-C}), we exclude accounts flagged as suspicious before the start of training (as they are more likely to be compromised in the future). Finally, to eliminate fakes, one of our classification exercises (\S\ref{section:results-CE-C}) attempts to eliminate all recently-created or dormant fakes by selecting for training only accounts that are older than two months old and have at least one month of activity (our assumption is that once fakes generate enough activity the LSOSP can detect them through existing techniques~\cite{yang:2011, wang:2013} as detecting fakes is easier than detecting compromised accounts~\cite{egele:2013}).

\begin{table}
    \footnotesize
    \centering
    \caption{Ground Truth Lag Experiment Results}
    \label{table:summary-lag-experiment}
    \begin{tabular}{|c|c|}

    \hline
    \textbf{Lag (in days)} & \textbf{Cumulative Percentage of Accounts (\%)} \\
    \hline
    
    1  & 74.31\% \\\hline
    2  & 78.90\% \\\hline
    3  & 81.98\% \\\hline
    4  & 84.89\% \\\hline
    5  & 86.50\% \\\hline
    6  & 88.59\% \\\hline
    \MyBGColorA 7  & \MyBGColorA 90.00\% \\\hline
    \MyBGColorA 21  & \MyBGColorA 98.59\% \\\hline
    28 & 100\% \\\hline
    
    \end{tabular}
    \vspace{-2em}
\end{table}

\vspace{+3pt}
\noindent
\textit{\textbf{Heuristics to reduce bias during classifier evaluation.}}
Our preliminary experiments suggest that user accounts that have been flagged as suspicious in the past are more likely to be flagged again in the future (a possible indication that their users are more vulnerable to attacks than the general user population). To provide a conservative (lower-bound) evaluation of the developed classifier's performance, we exclude all accounts that have been previously labeled as suspicious during training (i.e., flagged at any point during the training-LW or before). Moreover, one of our classification exercises (\S\ref{section:results-CE-C}), also excludes any accounts flagged as suspicious during the first month of the data collection. As a result, the developed classifier is evaluated on never seen before true positives.

\vspace{+3pt}
\noindent
\textit{\textbf{Heuristics to size the buffer window (BW).}}
It is expected that, at any LSOSP, detection of suspicious activity is not instantaneous, thus accounts may be under the control of an attacker for a while before they are flagged. We developed an experiment to estimate how aggressive is \companyshortname's suspicious activity flagging pipeline. 
For this experiment, we only rely on two types of events: flagging events for accounts marked as suspicious on day $d$ (extracted from set $S_d$) and login events for these accounts (extracted from set $L_d$). For this experiment we include only user accounts that have at least one login event and at least one flagging event within the period of time over which we run the experiment. We define the \textit{lag} per flagged user as the number of days between the first time that account is flagged and the most recent previous login event.

We run this experiment to gain some intuition for the lag of the flagging pipeline over a period of 30 days.
As shown in Table~\ref{table:summary-lag-experiment}, 90\% of accounts flagged within that period have a lag of at most one week and 98.6\% have a lag of less than three weeks. As such, we decided on a 1-week buffer window (BW) for most of our CEs, yet we also experiment with a 3-week BW (\S\ref{section:results-CE-D}). 

\section{Datasets}
\label{section:datasets}

Overall, we have access to 118 days ($\approx$4 months or $\approx$16 weeks) worth of production data collected from September 1st, 2016 to the December 27th, 2016. Overall, these datasets are representative of any LSOSP with a global user base, an extensive set of offered online services, as well as the latest techniques to identify potentially compromised accounts. 

\subsection{Extracting Features}

We have access to two datasets extracted and updated daily from production servers. The first is a dataset that includes features associated with all login events. Whenever a user logs-in to a service offered by \textit{\companyshortname} or has her session re-authenticated, a login event is recorded into this dataset with all relevant features that can be associated with the event at that time. We use this dataset to extract a minimal set of 13 basic and easy to compute features that reflect users' behavioral patterns (summarized in Table~\ref{table:summary-features}) from login traces at a day-level granularity, and then aggregate them for each user account as a way of characterizing its behavioral pattern over the DW.
It is important to note that we do not have access to any fine-grained account features such as account/user details. Importantly, we do not have access to any personally identifiable information. Moreover, given the diversity of the login methods as well as the services offered at \textit{\companyshortname}, the features extracted for each login event are not uniform and the set of features extracted for each user is sparse.

\begin{table}[t]
    \footnotesize
    \centering
    \caption{Summary of Low-Cost Features. \textmd{(from login traces)}}
    \label{table:summary-features}
    \setlength\tabcolsep{2pt}

    \begin{tabular}{|l|c|}
    
    \hline
    \textbf{Brief Description} & \textbf{Type} \\
    \hline

    \# Login Attempts & \multirow{14}{*}{Numeric} \\ \cline{1-1}
    
    \makecell[l]{\# Unique Login Sources (e.g., Web Login, Mobile Login, etc.)} &  \\ \cline{1-1}
    
    \makecell[l]{\# Unique Login Types (e.g., Password Login, Account Switch, etc.)} &  \\ \cline{1-1}

    \makecell[l]{\# Unique Login Statuses (e.g., Success, Session Extension, etc.)} &  \\ \cline{1-1}

    \makecell[l]{\# Unique Password Login Statuses (e.g., Success, Invalid Password, etc.)} &  \\ \cline{1-1}
    
    \makecell[l]{\# Unique Actions (e.g., Login/Logout, Device Authentication, etc.)} &  \\ \cline{1-1}

    \makecell[l]{\# Unique Login Geographical Locations} &  \\ \cline{1-1}
    
    \makecell[l]{\# Unique Login Geographical Location Statuses \\ (e.g., Neutral Location, White-listed Location, etc.)} &  \\ \cline{1-1}

    \# Unique Login Autonomous Systems (ASNs) &  \\ \cline{1-1}

    \makecell[l]{\# Unique Login User Agents (e.g., Browser, Mobile App, etc.)} &  \\ \cline{1-1}

    \# Successful Logins &  \\ \cline{1-1}

    \# Unsuccessful Logins &  \\ \hline

    User has a 'verified' mobile number & 2-Categorical \\ \hline

    \end{tabular}
    \vspace{-2em}
\end{table}

\subsection{Groundtruth: Suspicious Account Flagging}
\label{section:background-company-flagging-pipeline}

At \textit{\companyshortname}, a list of accounts flagged as suspicious is generated daily by combining information from various sources that include human content moderators, manual reports from internal teams, user reporting, in addition to automated systems employing heuristics  (that include clustering techniques to identify anomalies, and regression models to identify spammers).  We use this daily list of accounts flagged as suspicious as our ground truth: we extract labels for accounts indicating whether it has been flagged as suspicious or not during the training/testing intervals.

Accounts flagged as suspicious may include compromised accounts, as well as fake accounts. It is worth noting that, on the one hand, flagging is a result of undesired behavior exhibited by an account (e.g., sending spam), that is then subsequently detected either manually by moderators or automatically by some internal security system. On the other hand, this may be due to association or similarity with prior flagged suspicious accounts in terms of account behavioral characteristics (e.g., accounts logging in from a blacklisted autonomous system).

For this study, we had access to the daily list of accounts flagged as suspicious and a high-level description of the system. The detailed internals of the flagging pipeline were not available. As a consequence, we are neither able to distinguish between the different classes of suspicious accounts nor to identify the reason why a particular account had been flagged. We believe that, the lack of such fine-grained information poses limited threats to the validity of our findings: on the one side we have developed heuristics to exclude attacker-controlled accounts from training (see \S\ref{section:heuristics}), and, on the other side, at this point our machine learning model aims to provide only predictive power (will an account be flagged as suspicious?) rather than explanatory power (why will the account be flagged?). We extend this discussion in \S\ref{section:discussion}.

\section{Evaluation Results}
\label{section:results}

\textbf{\textit{The Objectives of our Classification Exercises.}} We present four of the classification exercises (CEs) carried out at \textit{\companyshortname} labeled $CE_A$, $CE_B$, $CE_C$, and $CE_D$ in Table~\ref{table:summary-classification-exercises}. The table outlines the Training and Testing intervals assigned to each CE and their respective Data Window (DW), Buffer Window (BW), and Label Window (LW). For each CE, we have a specific objective:

\begin{itemize}[noitemsep,topsep=0pt,leftmargin=*]
    \renewcommand{\labelitemi}{\scriptsize$\blacksquare$}

    \item $CE_A$: evaluating the feasibility of our proposed pipeline, its applicability at \textit{\companyshortname}, and optimizing hyperparameters.
    
    \item $CE_B$: testing the tuned model on new data to ensure that no overfitting occurred in $CE_A$.
    
    \item $CE_C$: investigating how the performance of our classifier changes when excluding accounts previously flagged as suspicious (higher chance to be flagged again) or accounts that have little previous activity (lower chance to include fakes).
    
    \item $CE_D$: evaluating the impact of more training data (longer data and label windows) and more aggressive exclusion of potentially not-yet-flagged attacker-controlled accounts (longer buffer window).
    
\end{itemize}

\begin{table*}[!ht]
    \centering
    \caption{Summary of Classification Exercises (CEs). \textmd{Notation: DW - Data Window, BW - BufferW, LW - LabelW, H - Prediction Horizon}}
    \label{table:summary-classification-exercises}
    \setlength\tabcolsep{8pt}

    \begin{tabular}{|c|c|c|c|c|c|c|c|c|c|c|c|c|c|c|c|c|}

    \hline
    \multirow{2}{*}{\textbf{CE}}
     & \multicolumn{16}{|c|}{\textbf{Week}} \\
    \hhline{*{1}{~}*{16}{-}}
     & 1 & 2 & 3 & 4 & 5 & 6 & 7 & 8 & 9 & 10 & 11 & 12 & 13 & 14 & 15 & 16 \\
    \hline\hline
    
    \multirow{2}{*}{\textbf{A}} & \multicolumn{3}{|c|}{\MyBGColorA Train} & \multicolumn{2}{|c|}{\MyBGColorB Test} & \multicolumn{11}{|c|}{\MyBGColorB Extended Test} \\
    \hhline{*{1}{~}*{16}{-}}
     & \MyBGColorA DW & \MyBGColorA BW & \MyBGColorA LW & \MyBGColorB DW & \MyBGColorB LW & \multicolumn{11}{|c|}{\MyBGColorB Extended LW \textit{(H = [7, 90] days)}}\\
    \hline\hline
    
    \multirow{2}{*}{\textbf{B}} & \multicolumn{8}{|c|}{Unused} & \multicolumn{3}{|c|}{\MyBGColorA Train} & \multicolumn{2}{|c|}{\MyBGColorB Test} &
    \multicolumn{3}{|c|}{\MyBGColorB Extended Test} \\
    \hhline{*{1}{~}*{16}{-}}
     & \multicolumn{8}{|c|}{Unused} & \MyBGColorA DW & \MyBGColorA BW & \MyBGColorA LW & \MyBGColorB DW & \MyBGColorB LW & \multicolumn{3}{|c|}{\MyBGColorB Extended LW}\\
    \hline\hline

    \multirow{2}{*}{\textbf{C}} &
    \multicolumn{4}{|c|}{\MyBGColorC Pre-Process} &
    \multicolumn{4}{|c|}{Unused} & \multicolumn{3}{|c|}{\MyBGColorA Train} & \multicolumn{2}{|c|}{\MyBGColorB Test} &
    \multicolumn{3}{|c|}{\MyBGColorB Extended Test} \\
    \hhline{*{1}{~}*{16}{-}}
     & \multicolumn{4}{|c|}{\MyBGColorC Pre-Process} & \multicolumn{4}{|c|}{Unused} & \MyBGColorA DW & \MyBGColorA BW & \MyBGColorA LW & \MyBGColorB DW & \MyBGColorB LW & \multicolumn{3}{|c|}{\MyBGColorB Extended LW}\\
    \hline\hline
    
    \multirow{2}{*}{\textbf{D}} &
    \multicolumn{9}{|c|}{\MyBGColorA Train} &
    \multicolumn{6}{|c|}{\MyBGColorB Test} & 
    \MyBGColorB Ext. Test \\
    \hhline{*{1}{~}*{16}{-}}
     & \multicolumn{3}{|c|}{\MyBGColorA DW} & \multicolumn{3}{|c|}{\MyBGColorA BW} & \multicolumn{3}{|c|}{\MyBGColorA LW} & \multicolumn{3}{|c|}{\MyBGColorB DW} & \multicolumn{3}{|c|}{\MyBGColorB LW} & \MyBGColorB Ext. LW\\
    \hline
    
    \end{tabular}
    \vspace{-1em}
\end{table*}

\begin{table*}[!hb]
    \centering
    \caption{Summary of Results using an Operating Threshold (T) = 0.5 for Different Prediction Horizons (H days).\textsuperscript{\ref{footnote:result_tables}}}
    \label{table:summary-results-threshold-5}
    \setlength\tabcolsep{5pt}
    \begin{tabular}{|c|c|c|c|c|c|c|c|c|c|c|c|c|c|c|}

    \hline
    \multirow{4}{*}{\textbf{CE}}
     & \multicolumn{14}{|c|}{\textbf{Performance Evaluation Metrics}} \\
    \hhline{*{1}{~}*{14}{-}}
     & \multirow{2}{*}{H$_{Min}$} & \multirow{2}{*}{H$_{Max}$} & 
     \multicolumn{2}{|c|}{H=H$_{Min}$}
     &  \multicolumn{4}{|c|}{\MyBGColorA H=7} & \multicolumn{2}{|c|}{\MyBGColorB H=21}
     & \multicolumn{2}{|c|}{\MyBGColorA H=30}
     & \multicolumn{2}{|c|}{\MyBGColorB H=H$_{Max}$} \\
    \hhline{*{3}{~}*{12}{-}}
     &  &  & AUC & BTR  
     & \MyBGColorA PRE & \MyBGColorA REC & \MyBGColorA ACC & \MyBGColorA FPR & \MyBGColorB PRE & \MyBGColorB REC & \MyBGColorA PRE & \MyBGColorA REC & \MyBGColorB PRE & \MyBGColorB REC \\
    \hline
    
    \textbf{A} & 7 & 90 & 0.928 & 85.61\% & \MyBGColorA \textbf{6.38\%} & \MyBGColorA \textbf{46.87\%} & \MyBGColorA 99.43\% & \MyBGColorA 0.52\% & \MyBGColorB \textbf{19.79\%} & \MyBGColorB 45.81\% & \MyBGColorA \textbf{20.14\%} & \MyBGColorA 43.81\% & \MyBGColorB \textbf{24.99\%} & \MyBGColorB 31.02\% \\
    \hline
    
    \textbf{B} & 7 & 30 & 0.910 & 82.14\% & \MyBGColorA 3.78\% & \MyBGColorA 41.26\% & \MyBGColorA \textbf{99.50\%} & \MyBGColorA \textbf{0.46\%} & \MyBGColorB 18.18\% & \MyBGColorB 46.82\% & \MyBGColorA 19.98\% & \MyBGColorA 42.28\% & \MyBGColorB 19.98\% & \MyBGColorB 42.28\% \\
    \hline
    
    \textbf{C} & 7 & 30 & 0.922 & 84.42\% & \MyBGColorA 3.18\% & \MyBGColorA 42.96\% & \MyBGColorA 99.38\% & \MyBGColorA 0.58\% & \MyBGColorB 16.58\% & \MyBGColorB 57.32\% & \MyBGColorA 18.33\% & \MyBGColorA \textbf{50.62\%} & \MyBGColorB 18.33\% & \MyBGColorB \textbf{50.62\%} \\
    \hline
    
    \textbf{D} & 21 & 34 & \textbf{0.947} & \textbf{89.41\%} & \multicolumn{4}{|c|}{\MyBGColorA H < H$_{Min}$} & \MyBGColorB 10.64\% & \MyBGColorB \textbf{57.42\%} & \MyBGColorA 11.68\% & \MyBGColorA 48.96\% & \MyBGColorB 12.34\% & \MyBGColorB 48.13\% \\
    \hline
    \end{tabular}
\end{table*}

\begin{table*}[!hb]
    \centering
    \caption{Summary of Results using an Operating Threshold (T) = 0.9 for Different Prediction Horizons (H days).\textsuperscript{\ref{footnote:result_tables}}}
    \label{table:summary-results-threshold-9}
    \setlength\tabcolsep{7pt}
    \begin{tabular}{|c|c|c|c|c|c|c|c|c|c|c|c|c|}

    \hline
    \multirow{4}{*}{\textbf{CE}}
     & \multicolumn{12}{|c|}{\textbf{Performance Evaluation Metrics}} \\
    \hhline{*{1}{~}*{12}{-}}
     & \multirow{2}{*}{H$_{Min}$} & \multirow{2}{*}{H$_{Max}$}
     & \multicolumn{2}{|c|}{H=H$_{Min}$}
     & \multicolumn{4}{|c|}{\MyBGColorA H=H$_{Min}$}
     & \multicolumn{4}{|c|}{\MyBGColorB H=H$_{Max}$}\\
    \hhline{*{3}{~}*{10}{-}}
     &  &  & AUC & BTR
     & \MyBGColorA PRE & \MyBGColorA REC & \MyBGColorA ACC & \MyBGColorA FPR & \MyBGColorB PRE & \MyBGColorB REC & \MyBGColorB ACC & \MyBGColorB FPR \\
    \hline
    
    \textbf{A} & 7 & 90 & 0.928 & 85.61\% & \MyBGColorA 12.92\% & \MyBGColorA ~0.47\% & \MyBGColorA \textbf{99.92\%} & \MyBGColorA \textbf{0.0024\%} & \MyBGColorB 33.99\% & \MyBGColorB ~0.20\% & \MyBGColorB 99.54\% & \MyBGColorB \textbf{0.0018\%} \\
    \hline
    
    \textbf{B} & 7 & 30 & 0.910 & 82.14\% & \MyBGColorA ~7.11\% & \MyBGColorA 13.15\% & \MyBGColorA 99.88\% & \MyBGColorA 0.0760\% & \MyBGColorB \textbf{35.96\%} & \MyBGColorB 12.90\% & \MyBGColorB 99.74\% & \MyBGColorB 0.0520\% \\
    \hline
    
    \textbf{C} & 7 & 30 & 0.922 & 84.42\% & \MyBGColorA ~6.91\% & \MyBGColorA \textbf{15.57\%} & \MyBGColorA 99.86\% & \MyBGColorA 0.0940\% & \MyBGColorB 35.33\% & \MyBGColorB \textbf{16.29\%} & \MyBGColorB 99.75\% & \MyBGColorB 0.0650\% \\
    \hline
    
    \textbf{D} & 21 & 34 & \textbf{0.947} & \textbf{89.41\%} & \MyBGColorA \textbf{26.19\%} & \MyBGColorA 14.45\% & \MyBGColorA 99.86\% & \MyBGColorA 0.0430\% & \MyBGColorB 28.47\% & \MyBGColorB 11.36\% & \MyBGColorB \textbf{99.82\%} & \MyBGColorB 0.0420\% \\
    \hline
    \end{tabular}
\end{table*}

\vspace{+3pt}
\noindent
\textit{\textbf{Summary of Results.}} 
Tables~\ref{table:summary-results-threshold-5} and~\ref{table:summary-results-threshold-9} summarize the results for all CEs carried out (their setup is outlined in Table~\ref{table:summary-classification-exercises}). For conciseness, we focus here only on the most relevant metrics
\footnote{\label{footnote:result_tables} Notation used in Tables~\ref{table:summary-results-threshold-5} and~\ref{table:summary-results-threshold-9}: AUC-Area Under Receiver Operating Characteristic Curve, BTR-\%-tile better than a random classifier, PRE-Precision, REC-Recall, ACC-Accuracy, FPR-False Positive Rate. Values in bold represent the best result for that performance metric.}
we collected. The two tables highlight how several metrics are impacted by the selected operating threshold $T$ of the classifier as well as by the duration of the prediction horizon (presented as Test-LW and Extended-Test-LW in Table~\ref{table:summary-classification-exercises} and whose combined size in days is denoted as \textit{the prediction horizon}: $H$). The tables present results for $T=0.5$ and $T=0.9$ and the values of $H=7, 21, 30, 34, 90~days$, in separate columns. Note that the minimum and maximum values of $H$ vary, depending on the CE.

In summary, these results show:

\begin{itemize}[noitemsep,topsep=0pt,leftmargin=*]
    \renewcommand{\labelitemi}{\scriptsize$\blacksquare$}

    \item High accuracy (ACC) $\approx$99.9\% and low false positive rate (FPR) <0.01\% for an operating threshold $T=0.9$,
    \item Good evidence for the absence of overfiting ($CE_B$ in \S\ref{section:results-CE-B}),
    \item Good balance between precision (PRE) and recall (REC): $\approx$18.33\% and $\approx$50.62\% respectively, when forecasting with a Horizon H = 30 days and Operating Threshold T = 0.5 ($CE_C$ in \S\ref{section:results-CE-C}),
    \item A small improvement after excluding recent/no activity accounts (more likely to be fakes) and those flagged as suspicious before training (Comparing $CE_B$ and $CE_C$),
    \item As the Horizon (H) increases, precision increases while recall stays roughly constant (We expand on this in \S\ref{section:horizon}),
    \item High AUC as shown in Figure~\ref{figure:results-roc} ($\approx$0.947\% for $CE_D$ in \S\ref{section:results-CE-D}), and
    \item More training data and a more aggressive exclusion of not-yet-flagged attacker-controlled accounts do not significantly impact classification performance ($CE_D$ in \S\ref{section:results-CE-D}).
\end{itemize}

\subsection{Pipeline Tuning (\boldmath$CE_A$)} 
\label{section:results-CE-A}

$CE_A$ uses data from the first five weeks of our traces for hyperparameter optimization. 
After a grid search, we extracted the hyperparameters of the best performing RF Classifier and fixed these for all the other CEs.

Our classifier performs well even for this heavily imbalanced dataset. The AUC attained is 0.928 that is 85.61\% better than the performance of a completely random classifier (with its AUC of 0.5). Note that at deployment, a LSOSP can tune the classifier to  achieve the desired balance between precision and recall by adjusting the classifier's operating threshold $T$ as highlighted by comparing the corresponding PRE/REC results in Tables~\ref{table:summary-results-threshold-5} and~\ref{table:summary-results-threshold-9}, and in Figure~\ref{figure:results-roc}. 

\textit{To pick only one datapoint from this experiment and highlight the performance of our classifier: more than one in four of the predicted accounts are actually later flagged to have suspicious activity within the next 90 days (PRE=24.99\%), and our classifier uncovers almost one third of the accounts later flagged as suspicious (REC=31.02\%) with a low false positive rate (FPR=0.42\%)}.

\subsection{Testing for Over-fitting (\boldmath$CE_B$)}
\label{section:results-CE-B}

To test for overfitting, through data reuse due to the hyperparameter optimization in $CE_A$, we repeated the same CE on entirely \textit{new data}, while using exactly the same experimental parameters. The model parameters identified during the hyperparameter optimization stage ($CE_A$) are reused for the rest of our CEs as well.  $CE_B$ uses the data collected over the last eight weeks of our traces: five weeks of data are used for training/testing similar to $CE_A$; this leaves three extra weeks of ground truth to explore the impact of a longer prediction horizon $H$ (in \S\ref{section:horizon}).

The classifier maintains the same levels of performance as in $CE_A$ (and these general trends are consistent across all CEs, none reusing $CE_A$ data). \textit{This suggests no overfitting for $CE_A$, and that the performance of the classifier is generally stable with different parameters (training/testing intervals, time windows, etc.).}

\subsection{Excluding New/Pre-Flagged Accounts (\boldmath$CE_C$)}
\label{section:results-CE-C}

Next, we investigate how the performance of our proposed classifier changes when we aggressively exclude accounts previously flagged as suspicious and accounts that have little past activity to increase the chance that we learn behaviours from accounts controlled only by legitimate users as explained in \S\ref{section:heuristics}. $CE_C$ excludes accounts that do not have any login activity within the one month pre-processing interval at the start of the experiment and any accounts that are flagged as suspicious within that same one month interval. With this extra pre-processing step, while there was a small improvement in performance, we did not observe any major impact.  \textit{This is an indication that fakes are unlikely to be a major problem at~\textit{\companyshortname}}. 

\subsection{Expanding the Training Data and BW (\boldmath$CE_D$)}
\label{section:results-CE-D}

Finally, as an extra experiment, we test our proposed classifier's performance when we increase the volume of training data, and \textit{aggressively increase the duration of the Buffer Window}. To this end, we decided to employ \textit{three week} windows (for DW, BW, LW during training/testing respectively) as opposed to \textit{one week}.
For $CE_D$, the proposed classifier attained the highest AUC among all our CEs $\approx$0.947 (89.41\% better than random). The classifier also achieved a PRE and REC of 10.64\% and 57.42\% respectively at $H$ = 21 days and $T$ = 0.5. At $T = 0.9$, precision reached 28.47\%, recall 11.3\% and the FPR an impressive 0.04\% (at $H=34$ days). 

\subsection{The Impact of the Prediction Horizon}
\label{section:horizon}
\begin{figure}
    \setlength{\belowcaptionskip}{-14pt}
    \includegraphics[height=1.65in]{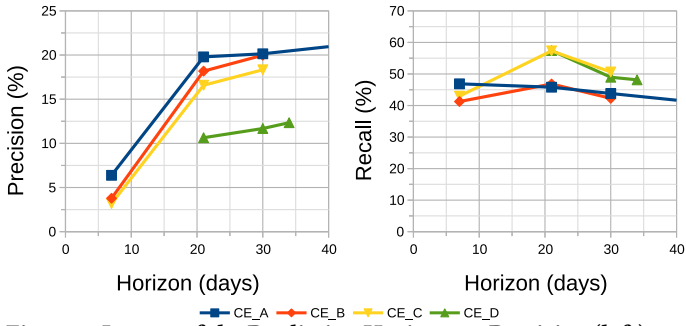}
    \caption{Impact of the Prediction Horizon on Precision (left) and Recall (right) at operating threshold $T=0.5$}
    \label{figure:results-pre-rec-horizon}
\end{figure}

Our classifier's precision markedly improves with the depth of the prediction horizon $H$ (Figure~\ref{figure:results-pre-rec-horizon}).
Some of the accounts that are false positives for a small precision window then become true positives as the prediction window increases. We speculate that those accounts are owned by users that do not have the ability or the interest to fend off social engineering attacks, and thus a longer horizon increases the chance that they fall victim to an attack, and then generate suspicious activity which gets them flagged during the longer prediction horizon.

We note that although precision increases with the prediction horizon, recall only stays stable (Figure~\ref{figure:results-pre-rec-horizon}). We hypothesize that the reason is that, during training, our classifier learns behavioral patterns from a discrete set of attacks only. As new attacks are developed, new categories of users become compromised and then their behavior is labeled as suspicious (yet our classifier can not label them as it has not been exposed to this data).

\subsection{Feature Importance}
\label{section:feture-importance}

Since we chose a Random Forest binary classifier, we can extract information regarding the relative importance of the features selected. For $CE_A$ the top four most important features are: the number of distinct geographic locations (20.07\%), the number of successful login events (15.94\%), the number of distinct ASNs (15.14\%), and the number of unsuccessful logins (8.697\%). While the relative importance of the features does not give an intuition as to why those features are suitable for the early identification of suspicious accounts in particular, they indicate that the same set of features that are typically used to flag suspicious accounts after the fact can also be used as early warning signs. We omit the relative feature importances for the rest of the CEs as they do not differ significantly from those presented for $CE_A$.

\section{Related Work}
\label{section:related-work}

Statistical methods (including ML) have achieved widespread adoption within LSOSPs not only to provide rich business features (e.g., product recommendations) but also for cybersecurity purposes. For instance, such approaches have been used for detecting compromised accounts, fake accounts, spam, and phishing. None of these approaches has focused on evaluating the feasibility of predicting which legitimate accounts are more vulnerable and likely to be compromised in the future (our long term aim). In this section each paragraph focuses on a specific area, surveys some related approaches, and outlines the statistical methods and features used.

\textbf{Compromised Accounts.} Egele et al.~\cite{egele:2013} combined statistical modeling and anomaly detection techniques in order to detect compromised accounts on Online Social Networks (OSNs). Their approach was based on identifying sudden changes in user behavioral patterns in addition to observing whether those changes are common to a large group of accounts therefore potentially a result of a malicious campaign. Thomas et al.~\cite{thomas:2014} employed clustering and classification (via logistic regression) in order to detect account hijacking on Twitter. Their approach was based on the observation that legitimate account owners frequently delete tweets posted via their accounts after recognizing the compromise. Those deletions are thus used as a feature to retroactively identify hijacked accounts and clustering is then used to detect similarly compromised accounts. Zhang et al.~\cite{zhang:2012} made use of a ML-based approach to automatically detect compromised accounts at a large academic institution. Their approach employed logistic regression on features extracted from web login and VPN authentication logs.

\textbf{Fake Accounts.} Yang et al.~\cite{yang:2011} proposed approaches to identify Sybil (i.e., fake) accounts on the Renren OSN. One approach was based on ML and employed Support Vector Machines (SVMs) on basic user-level features (e.g., the frequency of friendship requests and the fraction of accepted incoming friendship requests). Wang el al.~\cite{wang:2013} instead used clustering to identify fake accounts on Renren. Their approach clustered users with similar behavior based on features extracted from their clickstreams (e.g., the average session length, the average number of clicks per session).

\textbf{Spam.} Benevenuto et al.~\cite{benevenuto:2010} developed an ML-based approach to identify spammers on Twitter. Their approach was based on a non-linear Support Vector Machine (SVM) classifer with the Radial Basis Function (RBF) kernel and made use of both content- and user-level features (e.g., the age of the user account, the number of followers, the average number of URLs per tweet). Castillo et al.~\cite{castillo:2007} developed a ML-based approach using cost-sensitive decision trees to detect spam pages on the Web. Their approach makes use of content- and link-based features extracted from the Web graph (e.g., the ratio between the average degree of a page and that of its neighbours, number of words in the page/title). In the context of email spam, Blanzieri et al.~\cite{blanzieri:2008} carried out a survey of many of the approaches to detect email spam proposed in the literature based on statistical methods (including ML).

\textbf{Phishing.} Ludl et al.~\cite{ludl:2007} developed a ML-based approach to identify phishing web pages. Their approach was based on the C4.5 decision tree algorithm and made use of features extracted from a page's content as well as its URL (e.g., the number of forms/fields tags on the page, whether the page is served over HTTPS, whether the URL's domain appears on a Google whitelist). Whittaker et al.~\cite{whittaker:2010} developed a scalable ML-based approach to detect phishing websites that is used to maintain Google's phishing blacklist automatically. Their approach is based on a Random Forest (RF) classifier and employed both content-, host- and URL-based features (e.g., PageRank, the host geolocation/ASN).

\section{Summary and Discussion}
\label{section:discussion}

\textit{\textbf{Summary.}} We explore the feasibility of predicting the legitimate (i.e., not attacker-controlled) accounts more likely to generate suspicious activity in the future, a likely indication that they have fallen for a mass-scale social engineering attack. To this end, we propose an early warning system that employs supervised machine learning to identify the accounts whose behavioral patterns indicate that they are similar to other legitimate accounts that have been eventually labeled as suspicious in the past. We implement this early warning system at a Large-Scale Online Service Platform (LSOSP) and evaluate it on four months of real-world production data covering hundreds of millions of users. Our evaluation demonstrates that our approach is not only feasible but that it also offers promising classification performance based on which further defense mechanisms can be developed as we discuss below.

\vspace{+6pt}
\noindent
\textit{\textbf{Discussion.}} We continue by exploring several interrelated topics:

\vspace{+3pt}
\noindent
\textit{How can a defense system use information about which users are likely to be compromised in the future, and thus more `vulnerable', to enhance its robustness?}
User vulnerability can be thought of as an additional `signal' that can inform a number of defense mechanisms. For example, it can: \textit{(i)} serve as an indicator to prioritize the allocation of limited defense resources (e.g., use of human analyst time~\cite{ho:2017}, or compute-intensive filters~\cite{stein:2011}), \textit{(ii)} support differentiated defenses that take into account user vulnerability (e.g., additional CAPTCHAs on login attempts into vulnerable accounts~\cite{ahn:2003}, or imposing rate limits on the outbound messages of vulnerable users to slow-down the spread of multi-stage --- and potentially epidemic --- phishing attacks), \textit{(iii)} enable faster remediation of compromised accounts (e.g., by enabling more efficient inspection campaigns that focus on the accounts of vulnerable users instead of the entire user population~\cite{liu:2011}), \textit{(iv)} facilitate the detection of the origin of an attack (as, in effect, the differentiated response between vulnerable and robust users to similar interactions initiated by the same source can be used as a weak yet effective signal~\cite{boshmaf:2015}); and \textit{(v)} even facilitate the detection of new attacks (as, in effect, the differentiated response between vulnerable and robust --- yet otherwise similar --- user groups to the same `stimuli' is an indication of an attack). We explore the use of such information for several cybersecurity domains in~\cite{halawa:2016}.

\vspace{+3pt}
\noindent
\textit{Is the prediction quality good enough?}
Even if defense mechanisms based on vulnerability predictions can be imagined, an immediate subsequent question is whether the classification quality implied by our results (e.g., $PRE\approx15-25\%$, $REC\approx40-50\%$, and $FPR\approx0.1-0.5\%$) is good enough to support such mechanisms. While we have not yet extensively studied such mechanisms, our intuition is that this signal, although noisy, is useful. Consider, for example, defense resource prioritization - it is evident that a heuristic that uses this signal, as weak as it is, to prioritize resources is better than randomly allocating resources (when capacity is constrained).  Others have also experimented with a heuristic that harnesses the different response to similar requests between vulnerable and robust users~\cite{boshmaf:2015} to infer attack source(s) (although in the context of a social network). In this case, even a vulnerability predictor significantly weaker than the one we have obtained here has proven useful, leading to a technique that improves over the state-of-the-art. While the above indicates that even low quality predictions can still be used to improve defenses, we believe that the prediction quality threshold above which these mechanisms become valuable is context specific and we are studying this issue in a related project. 

\begin{figure}
    \setlength{\belowcaptionskip}{-14pt}
    \includegraphics[height=1.65in]{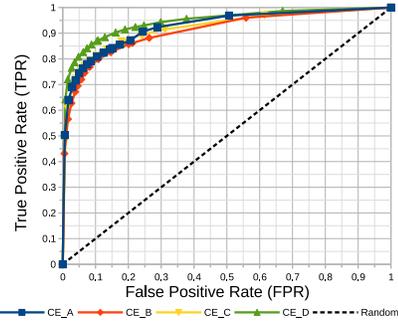}
    \caption{ROCs for all Classification Exercises.}
    \label{figure:results-roc}
\end{figure}

\vspace{+3pt}
\noindent
\textit{Why do we focus on minimizing the false positive rate (FPR)? What if the focus were on maximizing recall instead?}
We envisage that the predictions made by our early warning system will be used to better target existing defenses. As many of these defenses are not lightweight and may lead to increased friction for users (e.g., rate-limiting outbound emails of vulnerable users to prevent an attack outbreak, delaying incoming suspicious email addressed to vulnerable users to give enough time for more robust users to report mass-phishing emails), or allocating costly resources (e.g., human analyst time), the resulting cost of false positives is high: thus, we have focused on minimizing the FPR at the expense of lower recall. Other situations, however, offer a different cost/benefit balance between the false positive rate and recall. For these situations, our classifier can be tuned by either using lower threshold values ($T$ as highlighted by the ROC across all CEs available in Figure~\ref{figure:results-roc}), or by specifically optimizing for recall.

\vspace{+3pt}
\noindent
\textit{What are the threats to validity?}
Our study indicates that it is feasible to harness account behaviour to predict the accounts that are more likely to generate suspicious traffic in the future (an indicator that they may be compromised). There are two main concerns regarding the validity of our conclusions.
The first one relates to the quality of the ground truth we use --- this is a threat to validity common to any study using a methodology based on machine learning. 

The second one relates to the accuracy of the heuristics used to avoid learning behavioural patterns from accounts that may be controlled by an attacker (i.e., compromised or fake accounts) detailed in \S\ref{section:heuristics}. We prune: \textit{(i)} all accounts flagged for suspicious activity in the data window (DW) - as they are highly likely to be compromised, \textit{(ii)} all accounts flagged as suspicious in the buffer window (BW) - as these accounts are more likely to have been compromised but not yet flagged as such (thus contaminating our training data), \textit{(iii)} all accounts which have been labeled as suspicious at any point \textit{before} the training data window - as our experience shows that these accounts are more likely to be compromised again (in experiment $CE_C$); and, finally \textit{(iv)} new / low activity accounts (for which the system may not have enough history to determine whether the accounts are fakes). We run various experiments that compare the impact of these heuristics - even the most conservative experiments appear to support our conclusions.  

It is worth discussing, however, the alternative: assume that our heuristics fail to eliminate a large portion of attacker controlled accounts. Even in this case, we believe that our pipeline provides value through forecasting. Assume, for example, that these accounts are predominantly (dormant) fakes that mimic legitimate user behaviour.  In this case, our pipeline predicts the fakes that will likely be `awakened' by the attacker and start generating suspicious activity. Assume, on the other side, that these are compromised accounts not yet exploited by the attacker, then our pipeline predicts which compromised accounts are under the control of the attacker but not yet exploited. In this case as well the forecasting pipeline can give an early sign of the attacker resources and strategy.    

A final concern may be that our proposed approach may be learning the heuristics by which some accounts are flagged as suspicious in the ground truth (other accounts in the ground truth are flagged by humans). We believe that this represents a limited threat due to the way we formulated our forecasting problem (i.e., making future predictions) as opposed to the underlying heuristics which operate in real-time by design.

\vspace{+3pt}
\noindent
\textit{Why are the presented results positioned as lower-bounds?}
Our goal was to test the feasibility of our proposed approach within constraints related to access to data and computational resources. We believe that classification performance can be improved by: \textit{(i)} access to data beyond the login traces (e.g., email traces, or browsing patterns) which are likely to contain additional information that characterizes user behaviour; and \textit{(ii)} additional computational resources - as our exploration was  constrained by run-time feasibility in terms of data preprocessing (e.g., to extract complex aggregate features), model optimization, or sophisticated learning methods (e.g., deep neural networks) even though we used tens of computational nodes.